\documentclass[onecolumn,floatfix,superscriptaddress,showpacs,showkeys,nofootinbib,preprint]{revtex4}
\usepackage{epsfig}
\usepackage{amssymb,latexsym,amsmath}
\newcommand{\eq}[1]{\begin{align} #1 \end{align}}

\begin{document}

\title{Energy and Transverse Momentum Fluctuations \\in the Equilibrium Quantum Systems
       }

\author{M. I. Gorenstein}
 \affiliation{Bogolyubov Institute for Theoretical Physics, Kiev,
 Ukraine}
 \affiliation{Frankfurt Institute for Advanced Studies, Frankfurt,
 Germany}

\author{M. Rybczy\'nski}
\affiliation{Institute of Physics, Jan Kochanowski University, Kielce,
Poland}
\begin{abstract}
The fluctuations in the ideal quantum gases
are studied using the  strongly intensive measures  $\Delta[A,B]$ and
$\Sigma[A,B]$ defined in terms of two
extensive  quantities $A$ and $B$. In the present paper, these
extensive quantities are taken
as the motional variable, $A=X$,  the system
energy $E$ or transverse momentum $P_T$,  and number of
particles, $B=N$. This choice is most often considered
in studying the event-by-event
fluctuations and correlations in high
energy nucleus-nucleus collisions.
The recently proposed special normalization
ensures that $\Delta$ and $\Sigma$
%
are dimensionless and equal to unity for fluctuations given by the
independent particle model. In statistical mechanics,
the grand canonical ensemble
formulation  within the Boltzmann approximation
gives an example of independent particle model.
Our results demonstrate the effects
due to the Bose and Fermi statistics.
Estimates of the effects  of quantum statistics in
the hadron gas at temperatures and chemical potentials
typical for thermal models of hadron production in
high energy collisions
are presented.
In the case of massless particles and zero chemical potential
the  $\Delta$ and $\Sigma$ measures
are calculated analytically.
\end{abstract}



\maketitle
\section{Introduction}

Experimental and theoretical studies of the
event-by-event (e-by-e) fluctuations in
nucleus-nucleus (A+A) collisions give new information
about properties of the strongly interacting matter and its phases.
A possibility to observe signatures of the QCD matter critical point
inspired the energy and system size scan program of the
NA61/SHINE collaboration at the SPS CERN \cite{Ga:2009} and the
low energy scan program of the STAR and PHENIX collaborations at the
RHIC BNL \cite{RHIC-SCAN}.  In these studies one measures and then
compares the e-by-e fluctuations in  collisions of
different nuclei at different collision energies. The average
sizes of the created physical systems and their e-by-e
fluctuations are expected to be
rather different \cite{KGBG:2010}. This strongly
affect the observed hadron fluctuations, i.e. the measured
quantities would not describe the local physical properties of the
system but rather reflect the system size fluctuations. For
instance, A+A collisions with different centralities may produce a
system with  approximately the same local properties
(e.g., the same temperature and baryonic chemical potential)
but with the volume changing significantly from interaction to
interaction.  Note that in high energy collisions the
average volume  of created matter and its variations
from collision to collision are usually out of experimental control
(i.e. these volume variations are difficult or even
impossible to measure).

In the statistical mechanics the extensive quantity $A$
is proportional to the system
volume $V$, whereas  intensive quantity has a fixed finite value
in the thermodynamical limit $V\rightarrow \infty$.
The intensive quantities are used
to describe the local properties of a physical system. In
particular, an equation of state of the matter is usually
formulated in terms of the intensive physical quantities, e.g.,
the pressure is considered as a function of temperature and
chemical potentials.
In the statistical systems
outside of the phase transition regions,
a mean value of  fluctuating
extensive quantity,  $\langle A\rangle$, and its variance,
Var$(A)=\langle A^2 \rangle - \langle A \rangle^2$, are both
proportional to the volume $V$ in the limit of large volumes.
The scaled variance,
\eq{\label{omega-A}
\omega[A]~=~\frac{\langle A^2 \rangle~ -~ \langle A \rangle^2}{
\langle A \rangle}~,
}
is therefore an intensive quantity. However, the scaled
variance being an intensive quantity  depends on the system size
fluctuations.

Strongly intensive quantities introduced in Ref.~\cite{GG:2011} are
independent of the average volume and of
volume fluctuations. These quantities were suggested for and
are used in studies of e-by-e fluctuations of hadron
production in A+A collisions.
Strongly intensive measures of fluctuations
are defined in terms of two arbitrary
extensive quantities  $A$ and $B$.
In the present study we consider a pair of extensive variables --
the motional extensive variable $X=x_1+\dots x_N$ as a sum of single
particle variables $x_j$, with $j=1,\dots,N$, and  number
of particles $N$. These measures were recently
studied within the UrQMD simulations in Ref.~\cite{urqmd}.
The case of two hadron multiplicities
$A$ and $B$ in A+A collisions has been considered within the HSD
transport model in Ref.~\cite{HSD}. At the beginning we identify a single particle
variable $x$ with the particle energy $\epsilon$
and then consider the particle transverse momentum $p_T$.

The strongly intensive measure $\Delta[X,N]$ and $\Sigma[X,N]$
are defined as \cite{GG:2011}:
 \eq{\label{Delta-XN}
 &\Delta[X,N]
 ~=~ \frac{1}{C_{\Delta}} \Big[ \langle N\rangle\,
      \omega[X] ~-~\langle X\rangle\, \omega[N] \Big]~, \\
&\Sigma[X,N]
 ~=~ \frac{1}{C_{\Sigma}} \Big[ \langle N\rangle\,
      \omega[X] ~+~\langle X\rangle\, \omega[N] ~-~2\Big(\langle X\,N\rangle
      ~-~\langle X\rangle \langle N\rangle\Big)\Big]~,\label{Sigma-XN}
}
where $C_{\Delta}$ and $C_{\Sigma}$ are the normalization factors, and
the scaled variances $\omega[X]$ and $\omega[N]$  are given by
Eq.~(\ref{omega-A}).

In Ref.~\cite{GGP:2013} a special normalization for the strongly
intensive measures $\Delta$ and $\Sigma$ has been proposed.
Namely, the properly normalized strongly intensive quantities
assume the value {\it one} for fluctuations given by the independent
particle model (IPM). For the $X$ and $N$ extensive quantities the
proposed normalization reads \cite{GGP:2013}:
%
\eq{\label{norm}
 C_{\Delta}~=~C_{\Sigma}=~
\omega[x]\cdot \langle N\rangle~,~~~~~
 \omega[x]~\equiv~\frac{\overline{x^2}~-~\overline{x}^2}
{\overline{x}}~.
}
Note that the overline denotes averaging over a single particle
inclusive distribution, whereas $\langle \dots \rangle$ represents
averaging over multiparticle states of the system.

The first strongly intensive measure for fluctuations,
the so-called  $\Phi$ measure, was introduced
a long time ago in Ref.~\cite{GM:1992}.
The $\Phi$ quantity
for the ideal quantum gases was considered in Ref.~\cite{Mr:1998}.
There were numerous attempts  to use the
$\Phi$ measure describing fluctuations in
experimental data~\cite{Phi_data} and models~\cite{Phi_models}.
In general, however, $\Phi$
is a dimensional quantity and it does not have
a characteristic scale for a quantitative
analysis of e-by-e fluctuations
for different observables.
Note that the latter
properties  were clearly disturbing.
The $\Phi$ measure can be expressed in terms
of $\Sigma$~\cite{GG:2011}. A presence of additional
fluctuation measure $\Delta$ and utilization
of special normalization conditions for both $\Delta$ and $\Sigma$
give essential advantages in application to the data analysis
in A+A collisions.

In the  present paper we  study the strongly intensive measures
(\ref{Delta-XN}) and (\ref{Sigma-XN}) with normalization
factors (\ref{norm}) for the relativistic ideal quantum gases
in the grand canonical ensemble.
The paper is organized as follows. In Section~\ref{IQG} we
calculate the $\Delta[X,N]$ and $\Sigma[X,N]$
quantities for the ideal quantum gases
in the grand canonical ensemble. Analytical and numerical results
suitable for the hadron gas created in A+A
collisions are presented in Section~\ref{pion}.
A summary in Section~\ref{sum} closes the article. The calculation
details  are given in the Appendix.

\section{Ideal Quantum Gas}\label{IQG}

The grand canonical ensemble (GCE) partition function reads:
\eq{\label{Xi}
\Xi(V,T,\lambda)~=~\sum_N\sum_{\alpha}\lambda^N~\exp(-~\beta E_{\alpha})~,
}
where $V$ is the system volume,
$\beta \equiv T^{-1}$ is the inverse system temperature,
$\lambda\equiv \exp(\beta\mu)$ denotes the fugacity and $\mu$ the chemical potential.
The index $\alpha$ numerates the system quantum states, and $N$ is
the number of particles. The ensemble average values of the $k^{{\rm th}}$
moments ($k=1,2,\dots$) of any state quantity $A$ are calculated
as:
\eq{\label{A-av}
\langle A^k\rangle ~& =~ \frac{1}{\Xi}\sum_N
\sum_{\alpha}~A^k~\lambda^N\,\exp(-~\beta E_{\alpha})~.
 }
The GCE partition function (\ref{Xi}) can be presented in the form
\eq{\label{Xigce}
 \Xi~=~\exp\left\{~V~\eta^{-1}~d
\int\frac{d^3p}{(2\pi)^3}~\ln\left[1~+\eta~
\lambda~\exp(-\beta\,\epsilon)\right]\right\}~,
}
where $d$ is the number of particle internal degrees of freedom
and $\epsilon\equiv \sqrt{m^2+{\bf p}^2}$ is the particle energy
with $m$ being the particle mass and ${\bf p}$ its momentum. The
values $\eta=-1$ and $\eta=1$ correspond to the Bose and Fermi
statistics, respectively, whereas $\eta=0$ to the Boltzmann
approximation.
Using the presentation (\ref{Xigce}) one can calculate
the averages (\ref{A-av}) for the 1$^{{\rm st}}$ and 2$^{{\rm nd}}$ moments
of the energy $E$ and number of particles $N$:
 \eq{\label{N}
\langle N\rangle ~& =~ \frac{1}{\Xi}\lambda\frac{\partial
\Xi}{\partial \lambda}~=~V\rho~,~~~~   \rho  \equiv  d
\int\frac{d^3p}{(2\pi)^3}~
\frac{1}{\lambda^{-1}\exp(\epsilon/T)+\eta}~,
\\
%
%
%
\langle N^2\rangle  ~& = ~
\frac{1}{\Xi}\left(\lambda\frac{\partial}{\partial
 \lambda}\right)^2~\Xi~=~V^2\rho^2+
V\, I_N~,~~~~  I_N  \equiv
d\int\frac{d^3p}{(2\pi)^3}~\frac{\lambda^{-1}\exp(\epsilon/T)}
{\left[\lambda^{-1}\exp(\epsilon/T)+\eta\right]^2}~,
 \label{N2}
\\
%
%
%
%
\langle E\rangle ~ & =~
-~\frac{1}{\Xi}\frac{\partial\Xi}{\partial
\beta}~=~V\,\varepsilon~, ~~~~\varepsilon~ \equiv  ~
d~\int\frac{d^3p}{(2\pi)^3}~\frac{\epsilon}
{\lambda^{-1}\exp(\epsilon/T)+\eta}~,
}
\eq{
\langle E^2\rangle  ~& =~ \frac{1}{\Xi}\frac{\partial^2}{\partial
 \beta^2}\Xi~=~ \langle E\rangle^2\,+\,
V\,I_E~,~~~~ I_E  \equiv
d\int\frac{d^3p}{(2\pi)^3}~\frac{\epsilon^2\lambda^{-1}\exp(\epsilon/T)}
{\left[\lambda^{-1}\exp(\epsilon/T)+\eta\right]^2}~,
 \label{E2}\\
\langle EN\rangle
 ~ & =
-\frac{1}{\Xi}\frac{\partial}{\partial \beta}\lambda\frac{
\partial}{\partial \lambda}\Xi = \langle N\rangle \langle E\rangle \,+VI_{EN}~,
~~~~ I_{EN}  \equiv
d\int\frac{d^3p}{(2\pi)^3}~\frac{\epsilon\lambda^{-1}\exp(\epsilon/T)}
{\left[\lambda^{-1}\exp(\epsilon/T)~+~\eta\right]^2}~,
 \label{NE}
 }
where $\rho\equiv\langle N\rangle/V$ and $\varepsilon \equiv \langle E\rangle/V$ denote
the particle number density and the energy density, respectively.

From Eqs.~(\ref{N}-\ref{NE}) one finds
for the scaled variances:
\eq{\label{omegas}
\omega[N]~\equiv~\frac{\langle N^2\rangle~-~\langle
N\rangle^2}{\langle N\rangle}~=~\frac{I_N}{\rho}~,~~~~~~
\omega[E]~\equiv~\frac{\langle E^2\rangle~-~\langle E\rangle^2}{\langle E\rangle }
~=~\frac{I_E}{\varepsilon}~.
}
They describe the fluctuations of the number of particles and the
system energy at  fixed volume $V$. The scaled variances
in Eq.~(\ref{omegas}) are
intensive quantities,  they depend only on $T$ and $\mu$. The
quantities (\ref{omegas}) are independent of the particle
degeneracy factor $d$. Note that there is a (positive) correlation
between the energy $E$ and particle number $N$:
\eq{\label{EN-corr}
\langle EN\rangle ~-~\langle E\rangle ~\langle N\rangle ~=~V\,I_{EN}~>~0~.
}

The moments of single particle energy $\epsilon$ ($k=1,2$) are
\eq{\label{e}
\overline{\epsilon^k}~&=~ \frac{d}{\rho}
\int\frac{d^3p}{(2\pi)^3}~
\frac{\epsilon^k}{\lambda^{-1}\exp(\epsilon/T)+\eta}~,
}
and the scaled variance $\omega[\epsilon]$ is
\eq{\label{omega-eps}
\omega[\epsilon]~=~\frac{\overline{\epsilon^2}~-
~\overline{\epsilon}^2}{\overline{\epsilon}}~.
}

Calculating $\Delta[E,N]$ and $\Sigma[E,N]$ according to
Eqs.(\ref{Delta-XN}-\ref{norm}) one obtains:
\eq{
\label{D-GCE}
\Delta[E,N]~&=~\frac{1}{\omega[\epsilon]}~\Big[\omega[E]~-~\overline{\epsilon}
\cdot \omega[N]\Big]~=~\frac{1}{\omega[\epsilon]\cdot \rho}~\Big[\frac{I_E}{\overline{\epsilon}}~-~
\overline{\epsilon}\cdot I_N\Big]~,\\
\Sigma[E,N]~&=~\frac{1}{\omega[\epsilon]}~\Big[\omega[E]
~+~\overline{\epsilon} \cdot \omega[N]~-~2\,\frac{I_{EN}}{\rho}\Big]~
=~\frac{1}{\omega[\epsilon]\cdot \rho}~\Big[\frac{I_E}{\overline{\epsilon}}
~+~\overline{\epsilon}\cdot I_N~-~2\,I_{EN}\Big]~.\label{S-GCE}
}
Note that our choice of the normalization (\ref{norm}) makes
$\Delta[E,N]$ and $\Sigma[E,N]$ dimensionless. These quantities
are also independent of the degeneracy factor $d$.

The GCE within Boltzmann approximation satisfies the assumptions of
IPM. Thus, one expects for the Boltzmann gas
\eq{\label{SD-B}
\Delta^{{\rm Boltz}}[E,N]~=~\Sigma^{{\rm Boltz}}[E,N]~=~1~.
}
This can be easily proven, as for the $\eta=0$ one finds from
Eq.~(\ref{N}-\ref{NE}):
\eq{\label{SD-GCE}
\omega[N]=\frac{I_N}{\rho}=1~,~~~~\omega[E]=\frac{I_E}
{\overline{\epsilon}\cdot \rho}=~\frac{\overline{\epsilon^2}}{\overline{\epsilon}}
~,~~~~
\frac{\langle E\,N\rangle -\langle E\rangle\,\langle N\rangle}
{\langle N\rangle}= \frac{I_{EN}}{\rho}~=~\overline{\epsilon}~,
}
and Eqs.~(\ref{D-GCE}) and (\ref{S-GCE}) are transformed to
Eq.~(\ref{SD-B}). Using Eqs.~(\ref{A6}-\ref{A11}) from the Appendix
the following general relations can be proven
\eq{\label{BBF}
& \Delta^{{\rm Bose}}[E,N]~<~\Delta^{{\rm Boltz}}=1~<~\Delta^{{\rm Fermi}}[E,N]~,\\
& \Sigma^{{\rm Fermi}}[E,N]~<~\Sigma^{{\rm Boltz}}=1~<~\Sigma^{{\rm Bose}}[E,N]~,\label{FBB}
}
i.e.  Bose statistics makes $\Delta[E,N]$ to be
smaller and $\Sigma[E,N]$
larger than unity, whereas Fermi statistics works
in exactly opposite way.

The strongly intensive measures $\Delta$ and $\Sigma$ are
independent of the volume and of its fluctuations. This is valid
within the GCE,  when the temperature and chemical potentials are
volume independent. In a presence of volume fluctuations, the
second moments $\langle E^2\rangle$, $\langle N^2\rangle$, and
$\langle EN\rangle$  will include the terms proportional to
$\langle V^2\rangle$ which describe the contributions of the
volume fluctuations to the fluctuations of $E$ and $N$.  The full
averaging  will then include both the GCE averaging
(\ref{N}-\ref{NE}) at  fixed volume $V$ and an additional
averaging over the volume fluctuations. The scaled variances
(\ref{omegas}) do not depend on the average volume of the system,
i.e. they are intensive quantities. However, they do depend on the
volume fluctuations, i.e. the scaled variances are not strongly
intensive quantities.
On the other hand, the straightforward calculations demonstrate
\cite{GG:2011} that contributions from the volume fluctuations to
$\Delta$ and $\Sigma$ are cancelled out and
Eqs.~(\ref{D-GCE},\ref{S-GCE}) remain valid, i.e. $\Delta$ and
$\Sigma$ are indeed the strongly intensive measures in the GCE.

\section{Hadron Gas}\label{pion}
In this Section we consider the measures $\Delta$ and $\Sigma$ for
the hadron gas with thermodynamical parameters typical for the
thermal models of A+A collisions.
\subsection{Massless Particles}\label{m0}
We start from the system with $m=\mu=0$.
In the case,
the calculations
of quantities entering Eqs.~(\ref{N}-\ref{NE}) can be performed analytically.
Using Eqs.~(\ref{A2},\ref{A3}) one finds:
\eq{\label{n0}
\rho~&=~\frac{d}{2\pi^2}\int_0^{\infty}p^2dp\,\frac{1}{\exp(p/T)~\pm~1}~=~
\frac{d\,\zeta(3)}{\pi^2}\,{3/4 \choose 1}\,T^3~ \cong~d\,{0.091 \choose 0.122}\,T^3~,
\\
%
%
%
\overline{\epsilon}~&=
~\frac{d}{2\pi^2\,\rho}\int_0^{\infty}p^2dp\,\frac{p}{\exp(p/T)~\pm~1}~=~
\frac{3\,\zeta(4)}{\zeta(3)}\,{7/6 \choose 1}\,T~\cong~\,{ 3.152 \choose 2.701}~T~,\\
\overline{\epsilon^2}~&=~\frac{d}{2\pi^2\,\rho}\int_0^{\infty}p^2dp\,\frac{p^2}{\exp(p/T)~\pm~1}~=~
\frac{12\,\zeta(5)}{\zeta(3)}\,{5/4 \choose 1}\,T^2~\cong~{ 12.941 \choose 10.352}\,T^2
~,\\
I_N~&=~\frac{d}{2\pi^2}
\int_0^{\infty}p^2dp\,\frac{\exp(p/T)}{[\exp(p/T)~\pm~1]^2}~=~\frac{d\,\zeta(2)}{\pi^2}
\,{ 1/2 \choose 1}~T^3 \cong~ d~{0.083 \choose 0.167}~T^3~,\\
I_{EN}~&=~\frac{d}{2\pi^2}
\int_0^{\infty}p^2dp\,\frac{p\,\exp(p/T)}
{[\exp(p/T)~\pm~1]^2}~=~\frac{3\,d\,\zeta(3)}{\pi^2}\,
{3/4  \choose 1}\,T^4~\cong~ d\,{0.274 \choose 0.365}\, T^4~,\label{IEN0}\\
I_E~&=~\frac{d}{2\pi^2}
\int_0^{\infty}p^2dp\,\frac{p^2\,\exp(p/T)}
{[\exp(p/T)~\pm~1]^2}~=~\frac{12\,d\,\zeta(4)}{\pi^2}\,{7/8 \choose 1}\, T^5~
\cong~d\, { 1.151 \choose 1.316}\,T^5~, \label{E20}
}
where $\zeta(s)$ is the Riemann zeta function: $\zeta(2)= \pi^2/6\cong 1.645$,
$\zeta(3)\cong
1.202$, $\zeta(4)=~ \pi^4/90\cong 1.082$, $\zeta(5)\cong 1.037$. The upper case in
Eqs.~(\ref{n0}-\ref{E20}) and in equations below
corresponds to fermions ($\eta=+1$) and the lower
one to bosons ($\eta=-1$).
From these equations it follows:
\eq{\label{QS}
\omega[\epsilon]~
\cong~ {0.954 \choose 1.132}~T~,~~~~
\overline{\epsilon}\cdot \frac{I_N}{\rho}~\cong~{2.875 \choose 3.697}~T~,~~~~
\frac{I_E}{\rho\,\overline{\epsilon}}~= ~
4~T~,~~~~
\frac{I_{EN}}{\rho} ~= ~3~T~.
}
Finally, one obtains
\eq{\label{DS-m0}
\Delta[E,N]~\cong~{ 1.179 \choose 0.268}~,
~~~~~\Sigma[E,N]~\cong~{0.917 \choose 1.499}~.
}
The strongly intensive measures $\Delta[E,N]$ and $\Sigma[E,N]$
(\ref{DS-m0}) possess the values which are independent of $T$.
This is evident as the temperature is the only dimensional
variable for the system with $m=\mu=0$, and $\Delta$ and $\Sigma$
measures are dimensionless quantities due to our normalization.

In the Boltzmann approximation $\eta=0$, the integrals in
Eqs.~(\ref{n0}-\ref{E20})
are reduced to $\int_0^{\infty} x^k \exp(-x)=k!$
with $k=2,3,4$. One obtains
\eq{\label{Boltz}
\omega[\epsilon]~
=~T~,~~~~
\overline{\epsilon}\cdot \frac{I_N}{\rho}~=3\,T~,~~~~
\frac{I_E}{\rho\,\overline{\epsilon}}~= ~
4~T~,~~~~
\frac{I_{EN}}{\rho} ~= ~3\,T~,
}
and Eq.~(\ref{SD-B}) is satisfied.

\subsection{Pion Gas}

The pion gas corresponds
to the Bose
statistics ($\eta=-1$) and $m_\pi\cong 140$~MeV. We consider the
chemical equilibrium  pion gas ($\mu_\pi= 0$) and one example
of chemical non-equilibrium ($\mu_\pi=100~$MeV).
\begin{figure}[ht!]
\epsfig{file=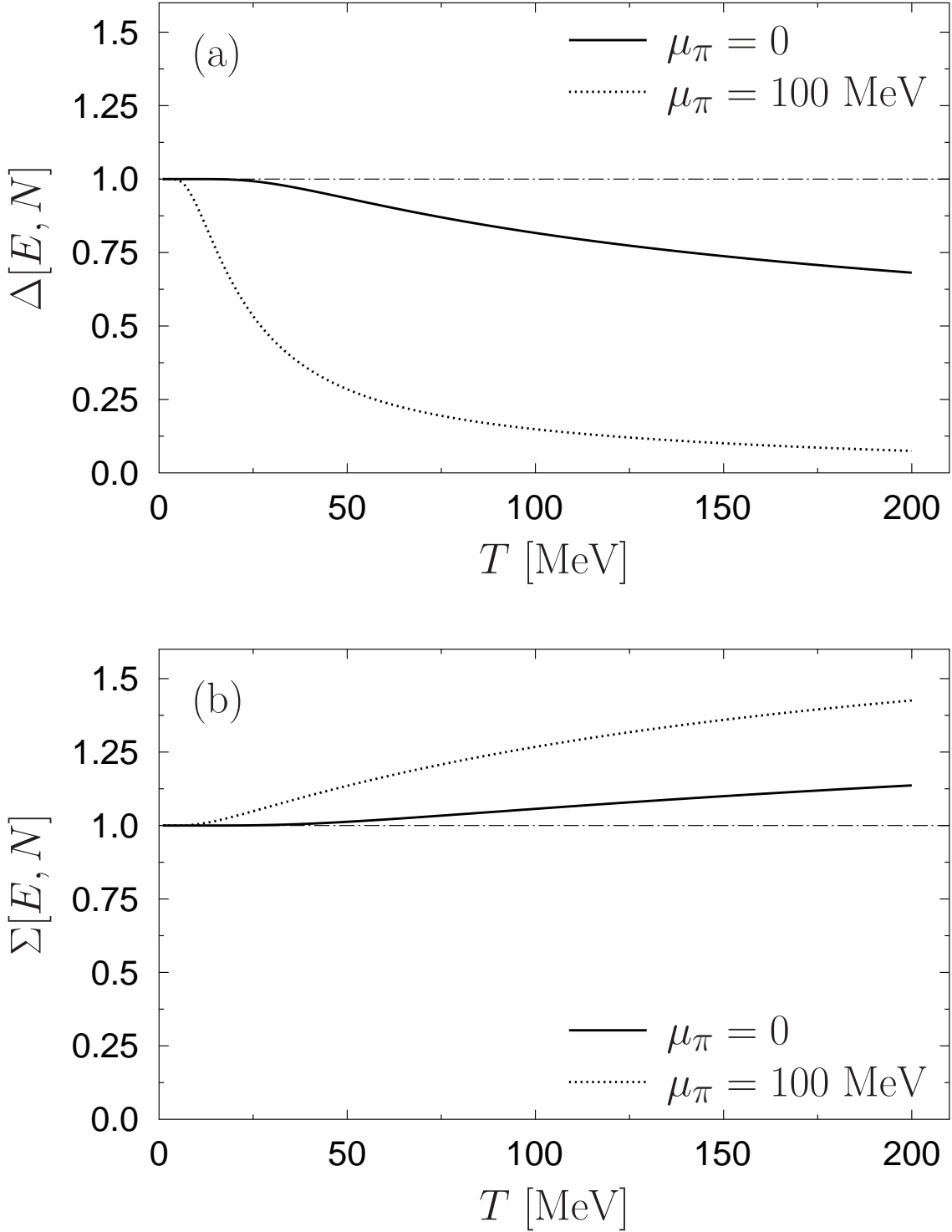,width=0.9\textwidth}
 \caption{The (a) $\Delta[E,N]$ and (b) $\Sigma[E,N]$
 for the pion gas as the functions of $T$. The solid lines correspond
to $\mu_\pi=0$ and dotted lines to $\mu_\pi=100$~MeV. The
horizontal dashed lines show the Boltzmann approximation
(\ref{SD-B}) equal to 1.
  \label{fig1} }
\end{figure}
Calculating the quantities $\rho$,  $\overline{\epsilon}$,
$\overline{\epsilon^2}$, $I_E$, $I_N$, $I_{EN}$ according to
Eqs.~(\ref{A6}-\ref{A11}) and $\omega[\epsilon]$ with
Eq.~(\ref{omega-eps}) one obtains $\Delta[E,N]$ by
Eq.~(\ref{D-GCE}) and $\Sigma[E,N]$ by Eq.~(\ref{S-GCE}). The
dependence $\Delta[E,N]$ and $\Sigma[E,N]$
on the temperature is shown in Fig.~\ref{fig1}. The two
lines are presented: the solid line for $\mu_\pi=0$ and the dotted
line for $\mu_\pi=100$~MeV. The horizontal line in
Fig.~\ref{fig1} (a) and (b) corresponds to the
Boltzmann approximation (\ref{SD-B}). This approximation
appears to be always valid at $T\ll m_\pi$.
In the ultra-relativistic
limit $T\gg m_\pi$ the results from Eq.~(\ref{DS-m0})
are approached, i.e approximately  0.27 for $\Delta[E,N]$
and 1.5 for $\Sigma[E,N]$. Note that solid and dotted
lines for $\Delta[E,N]$ verge towards their 
infinite temperature limit 0.27
from above and from below, respectively. Therefore, $\Delta[E,N]$
at $\mu_\pi=0$ has a minimum for an intermediate $T$ value.
However, the ultra-relativistic limit for pions
has of course only a mathematical
meaning, the hadron gas does not exist at $T>200$~MeV.

The typical
freeze-out temperatures in statistical and hydrodynamical models
of A+A collisions are $T=130\div 170$~MeV. In this temperature
region, the deviations of $\Delta[E,N]$ and $\Sigma[E,N]$ from the
IPM results (\ref{SD-B}) are quite significant, about 25\% and
10\%, respectively. These deviations are strongly enlarged for the
chemical non-equilibrium pion gas with $\mu_\pi> 0$.
The Bose statistics lead to the singular behavior of fluctuations
at $\mu_\pi\rightarrow m_\pi$, which corresponds to the Bose-Einstein
condensation of pions. We do not touch this problem in the present study.
For the fluctuations of pion multiplicity this was considered in
Ref.~\cite{BEC}.

In applications to A+A collisions one should however
take into account that a substantial fraction of the
final state pions come from the resonance decays.
These pions do not `feel' \cite{Mr:1998} the Bose statistics,
and thus the Bose statistics contribution to $\Delta[E,N]$ and $\Sigma[E,N]$
is reduced.

\begin{figure}[ht!]
\epsfig{file=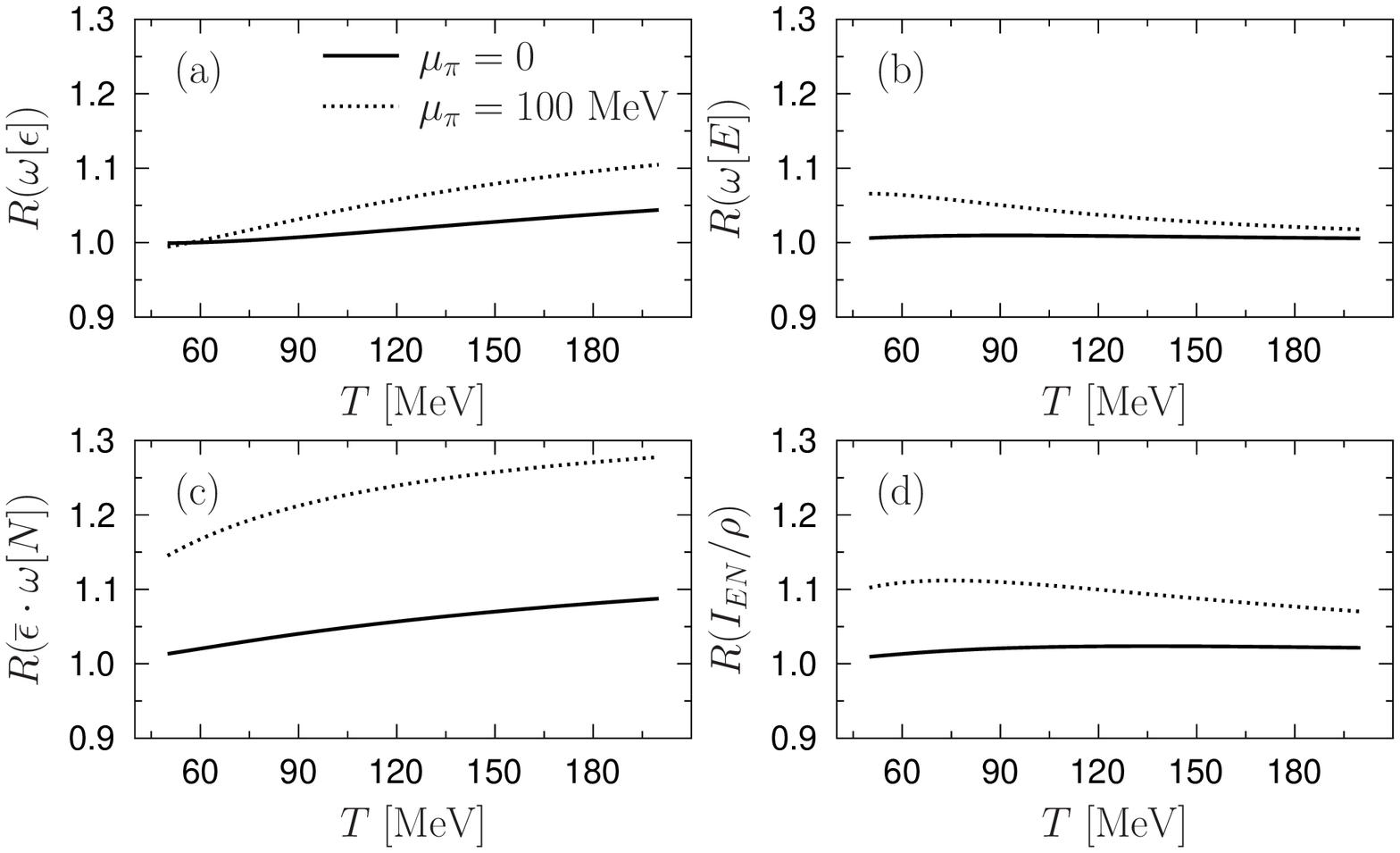,width=0.98\textwidth}
 \caption{The ratios (\ref{R})  for quantities
(a) $\omega[\epsilon]$, (b) $\omega[E]$, (c)
$\overline{\epsilon}\cdot
\omega[N]$, and (d) $I_{EN}/\rho$ in the pion gas
at $\mu_\pi=0$ (the solid lines) and $\mu_\pi=100$~MeV
(the dotted lines).  \label{fig2} }
\end{figure}

The strongly intensive measures $\Delta[E,N]$ (\ref{D-GCE}) and
$\Sigma[E,N]$ (\ref{S-GCE}) include the contributions from the
energy fluctuations $\omega[E]=I_E/(\rho\overline{\epsilon})$,
particle number fluctuations $\overline{\epsilon}\cdot\omega[N]=
\overline{\epsilon}\cdot I_N/\rho$, and correlations $I_{EN}/\rho$
between $E$ and $N$. The average particle energy can be
calculated as $\overline{\epsilon}=\langle E\rangle/\langle
N\rangle$. The suggested normalization requires also a knowledge
of $\overline{\epsilon^2}$ given by Eq.~(\ref{A8})  to find
the inclusive energy fluctuations $\omega[\epsilon]$
(\ref{omega-eps}).
Fore any physical quantity $A$ in the pion gas
we introduce the ratio
\eq{\label{R}
R(A)~=~\frac{A^{{\rm Bose}}}{A^{{\rm Boltz}}}~,
}
where $A^{{\rm Bose}}$ and $A^{{\rm Boltz}}$ are calculated
for  Bose statistics and within the Boltzmann approximation,
respectively. The  Bose and Boltzmann $A$-values will be
calculated at the same $T$ and $\mu_\pi$ values.
In Fig.~\ref{fig2} we show the ratios $R$ of the
pion gas quantities
$\omega[\epsilon]$, $\omega[E]$, $\overline{\epsilon}\cdot
\omega[N]$, and $I_{EN}/\rho$ to their Boltzmann approximations.
The later can be found by taking only the first terms
$n=1$ in the right-hand-sides of Eqs.~(\ref{A6}-\ref{A11}).
Deviations of the ratios in Fig.~\ref{fig2} from unity are
due to the Bose statistics effects for the corresponding physical
quantity. One observes that both $\omega[E]$ and $I_{EN}/\rho$ at
$\mu_{\pi}=0$ are approximately   insensitive  to   the   Bose
effects.  The Bose effects for $\Delta[E,N]$ and $\Sigma[E,N]$
seen in Fig.~\ref{fig1} are mostly due to the
quantum statistics contribution to $\overline{\epsilon}\cdot
\omega[N]$ and $\omega[\epsilon]$. Particularly, rather large values
of $R(\overline{\epsilon}\cdot
\omega[N])$ at $\mu_\pi=100$~MeV and large $T$
are seen in Fig.~\ref{fig2} (c).
Just these large values are responsible for a suppression of
$\Delta[E,N]$ and enhancement of $\Sigma[E,N]$
shown by dotted lines in Fig.~\ref{fig2} (a) and (b), respectively.
The above observation is also supported by
the analytical calculations at $m=\mu=0$
and becomes even stronger. A comparison
of Eq.~(\ref{QS}) and Eq.~(\ref{Boltz}) demonstrate that for
$m=\mu=0$ the values of $\omega[\epsilon]$ and
$\overline{\epsilon} I_N/\rho$ are sensitive to the effects of
quantum statistics whereas  $I_E/(\rho \overline{\epsilon})$ and
$I_{EN}/\rho$ are not.
%

The Bose statistics for the pion gas is the main source
of quantum statistics effects in the hadron gas with parameters
$T$ and $\mu$ typical for the hadron system created in A+A
collisions. The proton gas corresponds to
Fermi statistics ($\eta=1$) and $m=m_p\cong 938$~MeV.
The proton chemical potential is approximately equal to
the baryon chemical potential $\mu_B$ (additional contribution due
to electric chemical potential is negligible in high energy collisions).
The effects of quantum statistics in the proton gas increase with
increasing of both $T$ and $\mu_B$. However, the $T$ and $\mu_B$
values in the hadron gas are correlated: at small energies
of A+A collisions, large $\mu_B$ and small $T$ values appear,
whereas with increasing of collision energy $T$ moves to its maximum
of about 170~MeV and simultaneously $\mu_B$ approaches to zero.
For typical $T$ and $\mu_B$ values we find for the proton gas
\eq{\label{proton}
& \Delta[E,N]~\cong ~1.030~,~~~~ \Sigma[E,N]~\cong~0.997~,~~~~{\rm at}~~
T=150~{\rm MeV}~, ~~~ \mu_B=300~{\rm MeV}~,  \\
& \Delta[E,N]~\cong ~1.040~,~~~~ \Sigma[E,N]~\cong~0.997~,~~~~{\rm at}~~
T=100~{\rm MeV}~, ~~~ \mu_B=500~{\rm MeV}~.
}
A deviation of the $\Delta$ and $\Sigma$ quantities from 1 in an
ideal gas within  GCE is due to the
quantum statistics.
From Eq.~(\ref{proton}) one concludes that Fermi statistics effects
for the proton gas are quite small for typical $T$ and $\mu_B$ values
in the hadron gas: they give only a few percent contribution to $\Delta[E,N]$
and almost negligible contribution to $\Sigma[E,N]$.

\subsection{Transverse Momentum Fluctuations}
Using the above equations one can easily calculate the
measures $\Delta[P_T,N]$ and $\Sigma[P_T,N]$ for the transverse
momentum fluctuations. Since $p_T=p\cdot \sin(\theta)$ with
$p=|{\bf p}|$ and $\theta$ being the angle between the `beam'
$z$-axis, one gets for an arbitrary $f(p)$ function:
\eq{\label{pTp}
 \int d^3p ~p_T~f(p)~=~\frac{\pi}{4}\int d^3p~p~f(p)~,~~~~
 \int d^3p ~p^2_T~f(p)~=~\frac{2}{3}\int d^3p~p^2~f(p)~.
}

First, one needs to calculate the integrals (\ref{A6}-\ref{A11})
for $P$ and $N$ quantities, i.e. with $p$ and $p^2$
instead of $\sqrt{p^2+m^2}$ and $p^2+m^2$,
respectively. The integrals (\ref{A6}) for $\rho$ and (\ref{A9})
for $I_N$ remain unchanged. There are two new integrals:
\eq{
\overline{p}~&=~\frac{d}{2\pi^2\,\rho}\,\int_0^{\infty}p^2dp~\frac{p}
{\exp[\sqrt{p^2+m^2}/T]~-~1}~,\\
I_{PN}~&=~\frac{d}{2\pi^2} \int_0^{\infty}p^2dp\,
\frac{p~\exp[\sqrt{p^2+m^2}/T]}
{\Big[\exp (\sqrt{p^2+m^2}/T)~-~1\,\Big]^2}~,
}
instead of $\overline{\epsilon}$ and $I_{EN}$, respectively.
To calculate $\overline{p^2}$ and $I_P$
the following relations can be used:
\eq{
\overline{p^2}~=~\overline{\epsilon^2}~-~m^2~,~~~~ I_P~=~I_E~-~m^2\cdot I_N~.
}

\begin{figure}[ht!]
\epsfig{file=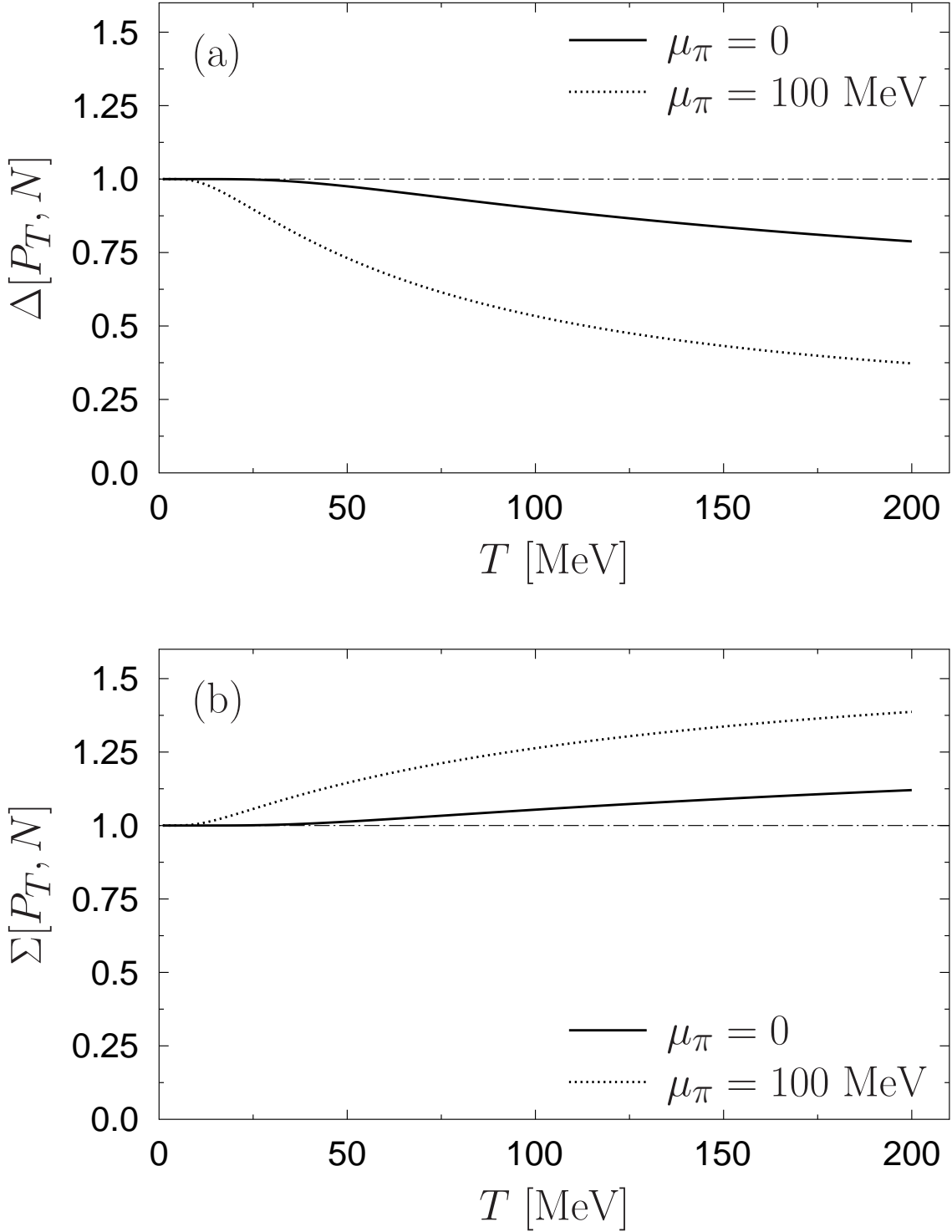,width=0.9\textwidth}
 \caption{The (a) $\Delta[P_T,N]$ and (b) $\Sigma[P_T,N]$
 for the pion gas as the functions of $T$. The solid lines correspond
to $\mu_\pi=0$ and dotted lines to $\mu_\pi=100$~MeV.
The
horizontal dashed lines show the Boltzmann approximation
(\ref{SD-B}) equal to 1.  \label{fig3} }
\end{figure}

Using Eq.~(\ref{pTp}) one then finds:
%
\eq{\label{p}
& \overline{p_T}~=~\frac{\pi}{4}\cdot \overline{p}
~,~~~~
\overline{p_T^2}~=~ \frac{2}{3}\cdot \overline{p^2}
~,~~~~
 \omega[p_T]~=~\frac{\overline{p_T^2}~-~\overline{p_T}^2}{\overline{p_T}}
~,\\
%
%
& I_{P_T}~= ~
\frac{2}{3}\cdot I_E
~,~~~~
I_{P_TN} ~= ~\frac{\pi}{4}\cdot I_{EN}~.
}

The results for $\Delta[P_T,N]$ and $\Sigma[P_T,N]$ in the pion
gas are shown in Fig~\ref{fig3}. From a comparison of this figure
with Fig.~\ref{fig1} one concludes that qualitative behavior of
$\Delta$ and $\Sigma$ measures for extensive quantities $[P_T,N]$
is the same as for $[E,N]$ ones. Quantitatively, the Bose effects
in Fig.~\ref{fig3} are smaller than the corresponding effects seen
in Fig.~\ref{fig1}.
At $m=\mu=0$ one obtains
\eq{\label{PT0}
\Delta[P_T,N]~\cong~{ 1.125 \choose
0.433}~,~~~~~\Sigma[P_T,N]~\cong~{0.931 \choose 1.398}~.
}
These values can be compared with the corresponding results
for $E$ and $N$,
Eq.~(\ref{DS-m0}).

\subsection{Connection to the $\Phi$ Measure}
The well-known fluctuation measure $\Phi$ was introduced in
Ref.~\cite{GM:1992}.
In a general
case, when $X=\sum_{i=1}^N x_i$ represents any motional extensive
quantity as a sum of single particle quantities,
one gets \cite{GG:2011}:
\begin{equation}\label{phi_x}
\Phi_X~=~\left[ \frac{\omega[x]\,\langle X \rangle} {\langle
N\rangle
}~\Sigma[X,N]\right]^{1/2}~-~\left[\overline{x^2}~-~\overline{x}^2\right]^{1/2}~,
\end{equation}
where $\Sigma[X,N]$ is given by Eq.~(\ref{Sigma-XN}) and
$C_{\Sigma}$ by Eq.~(\ref{norm}). Therefore, the $\Phi$ quantity can be expressed
via measure $\Sigma$.
%
At $m=\mu=0$ it then follows:
\eq{\label{Phi}
\Phi_E~&=~\Big[\overline{\epsilon}\cdot\omega[\epsilon]\Big]^{1/2}\,
\Big[\Big(\Sigma[E,N]\Big)^{1/2}~-~1\Big]~ =~{-\,0.074 \choose 0.392}~T~,\\
%
\Phi_{P_T}~&=~\Big[\overline{p_T}\cdot\omega[p_T]\Big]^{1/2}\,
\Big[\Big(\Sigma[P_T,N]\Big)^{1/2}~-~1\Big]~
=~{-\,0.056 \choose 0.283}~T~.
}
These results are in agreement with those obtained in
Ref.~\cite{Mr:1998}.

\section{Summary}\label{sum}
The  strongly intensive fluctuation measures  $\Delta$ and
$\Sigma$ have been studied  for the ideal Bose and Fermi gases
within the grand canonical ensemble.
In the present paper, the $\Delta$ and $\Sigma$ quantities
are considered for two specific
extensive  quantities  -- motional variable $X$ (either  the system
energy $E$ or transverse momentum $P_T$)  and number of
particles $N$. We have used
the normalization
of the strongly intensive measures
which makes them dimensionless and
equal to unity for fluctuations given by the
independent particle model. The grand canonical ensemble
within the Boltzmann approximation satisfies the conditions
of independent particle model.
Our results demonstrate deviations from the independent particle
model due to the Bose and Fermi statistics. We present
estimates of these quantum statistics effects for the hadron gas
with thermodynamical parameters typical for the thermal models of A+A
collisions. In the case of massless particles and zero chemical potential
the  $\Delta$ and $\Sigma$ measures
are calculated analytically. Numerical estimates for the Bose effects in
the pion gas at the temperatures from 0 to 200 MeV
are presented. For the Fermi gas of protons the quantum effects appear
to be quite small.

The measures $\Delta$ and $\Sigma$ are used
to study the event-by-event
fluctuations and correlations in high energy nucleus-nucleus and
proton-proton collisions. From our results it follows that
the Bose effects in the pion gas can be an important source
of the transverse momentum fluctuations, especially in chemically
non-equilibrium case with $\mu_{\pi}>0$.
However, other sources of dynamical fluctuations
and correlations (e.g., exact conservation laws within micro-canonical
ensemble, resonance decays, transverse collective flow,
fluctuations of temperature, correlations between temperature and
particle multiplicity, etc.)
should be considered to make a realistic comparison of
theoretical models with the data.

\vspace{0.3 cm}
\begin{acknowledgments}
We would like to thank Marek Ga\'zdzicki
and Stanislaw Mr\'owczy\'nski for
fruitful discussions and comments.
The work of M.I.G. was supported by the Program of
Fundamental Research of the Department of Physics and
Astronomy of NAS, Ukraine. This research was supported
by the National Science Center (NCN) under contract - 2011/03/B/ST2/02617.
\end{acknowledgments}

\appendix

\section{}\label{A2}
Eqs.~(\ref{n0}-\ref{E20}) for $m=\mu=0$
are reduced to the following integrals ($k=2,3,4$):
\eq{
%
%
& \int_0^{\infty}dx~\frac{x^{k}}
{\exp(x)~+~\eta}~=~k!\,\zeta(k+1)\,{1-2^{-k} \choose 1}~\label{A2},\\
& \int_0^{\infty}dx~\frac{x^{k}\exp(x)}{[\exp(x)~~+
~\eta]^2}~=~k!\,\zeta(k)\, { 1-2^{1-k} \choose 1}~, \label{A3}
}
where the upper case corresponds to $\eta=1$ (fermions) and lower one
to $\eta=-1$ (bosons). For $\eta=0$ (Boltzmann approximation)
integrals (\ref{A2}) and (\ref{A3}) become identical and equal to $k!$.
%
%
%
%
%
%
%

Using the series expansions,
\eq{
\frac{1}{\exp(z)+\eta}=\sum_{n=1}^{\infty}(-\eta)^{n-1} \exp(-\,n
z)~,~~~~\frac{\exp(z)}{[\exp(z)+\eta]^2}=\sum_{n=1}^{\infty}
(-\eta)^{n-1}\, n\,\exp(-\,n z)~,\label{A4}
}
one obtains ($y \equiv m/T$):
\eq{
&\rho~=~\frac{d}{2\pi^2}\int_0^{\infty}p^2dp\,
\frac{1}{\exp\Big[(\sqrt{p^2+m^2}-\mu)/T\Big]~+~\eta}~\nonumber \\
&=~
\frac{dT^3}{\pi^2}~y^2\,\sum_{n=1}^{\infty}
\frac{(-\eta)^{n-1}}{n}\,K_2(n\,y)\,
\exp\Big(\frac{n\,\mu}{T}\Big)~,\label{A6}
\\
%
%
& \overline{\epsilon}~=~\frac{d}{2\pi^2\,\rho}
\int_0^{\infty} p^2dp\, \frac{\sqrt{p^2+m^2}}{\exp \Big[(\sqrt{p^2+m^2}-\mu)/T \Big]~+~\eta }~\nonumber \\
&=~
\frac{d\,T^4}{16\pi^2\,\rho }\,y^4\,\sum_{n=1}^{\infty} (-\eta)^{n-1}\,
\Big[K_4(n\,y)~-~K_0(n\,y)\Big]
\,\exp\Big(\frac{n\,\mu}{T}\Big)~, \label{A7}\\
%
%
%
%
&\overline{\epsilon^2}~=~\frac{d}{2\pi^2\,\rho}
\int_0^{\infty}p^2dp\, \frac{p^2+m^2}{\exp\Big[(\sqrt{p^2+m^2}-\mu)/T\Big]~+~\eta}~\nonumber \\
&=~
\frac{d\, T^5}{32\pi^2}\,y^5\,\sum_{n=1}^{\infty} (-\eta)^{n-1}\,
\Big[K_5(n\,y)~+~K_3(n\,y)
-~2\,K_1(n\,m)\Big]
\,\exp\Big(\frac{n\,\mu}{T}\Big)~, \label{A8}
\\
%
%
%
%
&I_N~=~\frac{d}{2\pi^2} \int_0^{\infty}p^2dp\, \frac{\exp[(\sqrt{p^2+m^2}-\mu)/T]}
{\Big[\exp [(\sqrt{p^2+m^2}-\mu)/T]+\eta\Big]^2}
\nonumber \\
&=~\frac{d\,T^3}{2\pi^2}\,y^2\,\sum_{n=1}^{\infty}
(-\eta)^{n-1}\,K_2(n\,y)\,
\exp\Big(\frac{n\,\mu}{T}\Big)~,\label{A9}
}
\eq{
&I_{EN}~=~\frac{d}{2\pi^2} \int_0^{\infty}p^2dp\,
\frac{\sqrt{p^2+m^2}\,\exp[(\sqrt{p^2+m^2}-\mu)/T]}
{\Big[\exp [(\sqrt{p^2+m^2}-\mu)/T]~+~\eta\Big]^2}~\nonumber
\\
%
%
&=~
\frac{d\,T^4}{16\pi^2}\,y^4~\sum_{n=1}^{\infty}
(-\eta)^{n-1}\,n\,\Big[K_4(n\,y)~-~K_0(n\,y)\Big]\,
\exp\Big(\frac{n\,\mu}{T}\Big)~,\label{A10} \\
%
%
%
&I_E~=~\frac{d}{2\pi^2} \int_0^{\infty}p^2dp\,
\frac{(p^2+m^2)~\exp[(\sqrt{p^2+m^2}-\mu)/T]}
{\Big[\exp [(\sqrt{p^2+m^2}-\mu)/T]~+~\eta\Big]^2}~\nonumber \\
&=~
\frac{d\,T^5}{32\pi^2}\,y^5\,\sum_{n=1}^{\infty}
(-\eta)^{n-1}\,n\,\Big[K_5(n\,y)~+~
K_3(n\,y)~-~2K_1(n\,y)\Big]\,
\exp\Big(\frac{n\,\mu}{T}\Big)~,\label{A11}
}
where $K_l(z)$ are the modified Bessel functions.

\end{document}